\documentclass[ChapterTOCs]{book}	
	
\usepackage[sevenbyten]{centennial}
\usepackage{times,mathptmx}
\usepackage{graphicx}
\usepackage{rotating}
\usepackage{fancybox}
\usepackage{wrapfig}
\title{Nonlinear Optics}
\author{Peter Body}

\begin{document}

\frontmatter

\pagestyle{myheadings}
\tableofcontents
\mainmatter

\include{TestChap}

\markboth{Superconductivity}{A modern, but way too short history of  the theory of superconductivity at a high temperature.} 
\chapterauthor{Jan Zaanen}{Instituut-Lorentz for Theoretical Physics, Leiden University, Leiden, The Netherlands} 
\chapter{2.4 A modern, but way too short history of the theory of superconductivity at a high temperature.} 
\label{Chapter2.4} 
\section{Introduction} 
\label{1sec:Introduction} 
\secauthor{J. Zaanen}{Leiden University, The Netherlands} 
One cannot write the history of a war when it is still raging and this is certainly applying to the task we
 are facing dealing with the theory of superconductivity in the post BCS  era. 
The BCS theory was of course a
monumental achievement that deserves to be counted among the greatest triumphs
in physics of the twentieth century. With the discovery of high-{\em T}$_c$ superconductivity in
the cuprates in 1986\footnote{K.A. M\"uller, G. Bednorz, Z. Phys. B64 (1986) 189.} a consensus emerged
immediately that something else was at work than the classic (i.e. phonon driven) BCS mechanism. 
A quarter of a century later this subject is still contentious, perhaps even more so than at any
instance in the past.  It is plainly impossible to write an 'objective' history of any kind and this piece has no
pretense in this regard. I thought it would be useful to write instead  some sort of eyewitness account,
serving two potential readerships: in the first place the newcomers who wonder about the prehistory 
of some standard notions that appear as established and controversial at the same time.
The other customers are right now no more than a potentiality.  Imagine that the theory of 
high-{\em T}$_c$ superconductivity  will turn truly glorious at some point in the future. The era that I am discussing here
will then have a similar status as the 'dark ages' that preceded the discovery of BCS theory as described elsewhere 
in this volume.  I can imagine that an eyewitness account like the present one might be entertaining for the 
professional historians who want to chronicle this triumph of 21-th century physics. 

I am aware that this exposition is far from complete -- there are quite a number of ideas that are very interesting by
themselves, while they might be on the long run more consequential than the matters I discuss here.  However, 
given the length restrictions I just have not the space to give a truly comprehensive overview of everything that
has been explored. I have to restrict myself to a sketch of the history behind the notions that are right now most
on the foreground. Similarly, I do not claim rigor with regard to distributing the credits to the various individuals
involved. Although at some instances I will draw attention to contributions that I find are underexposed, this material 
is surely not intended as  background material for award committees.  

This eyewitness account is also colored with regard to the substances that are discussed.  
Whether we like it or not, humanity is a religious species and belief
systems play a big role when science is still in the making. One better be aware of it and in physics I like to
be myself a 'denomination tourist'. I have dwelled around among the various belief systems, getting into it to a degree 
that I felt myself to be a believer, to hop at that point to a next system. In the course of time I have settled into 
a faith that is right now quite popular: the  'agnostic denomination in high-{\em T}$_c$ 
superconductivity'.  Their dogma is  as follows: "the research in unconventional superconductivity is flourishing.  It is however
owned by the experimentalists. Energized by the high-{\em T}$_c$  puzzle, the experimentation in quantum 
matter leaped forward in the last twenty years, due to a variety of highly innovative instrumental developments. These new experimental facts  
have however just deepened the mystery for the theorists.  The agnostic zeal is anchored in the conviction that 
these fantastic new data are begging for an explanation in terms of mathematical equations of a beauty and elegance that will 
measure up to the  products of Einstein and Dirac. Right now we have not the faintest idea what these equations 
are but that should not keep us away from attempting to find them." 

Surely this has imprinted on the organization of this text. I will discuss matters in a sequence that is in 
a historical order (from old to new), while at the same time the sense of  the awareness of dealing with a 
mystery (the agnostic dogma) is increasing. 

The point of departure is the highly influential 'denomination' that rests on the believe that the theoretical framework 
as emerged in the 1950's in essence still suffices to explain the physics of the cuprates and so forth. The normal state
is a Fermi-liquid that is perhaps a bit obscured by perturbative processes. However, the essence of BCS theory is still fully intact and  
to explain the high-{\em T}$_c$  superconductivity one needs a 'superglue' that has more punch than the traditional phonons.
This superglue is than supposed to be embodied by spin fluctuations.  Since this 'fifties paradigm' has been around
for a long time it has turned into a framework that pretends to be quite qualitative. This  "holy trinity"
( Local Density Approximation (LDA) band structure, Random Phase Approximation (RPA)
 spin fluctuation superglue,  Eliasberg equations for the superconductivity)
is right now in the middle of a revival in the context of the superconductivity in the iron pnictides. It surely 
should be taken quite seriously but more than anywhere else the danger of religious illusions is around the corner. 
In this regard I perceive it as beneficial to be aware of the long, and contentious history of the spin fluctuation 
superglue idea. This is the narrative of section 2.4.2.

Although it has been a bit on the retreat recently, the next influential school can be called 'RVB, and related phenomena'. This refers to a
highly   imaginative set of ideas, that emerged in the wild early years of high-{\em T}$_c$  superconductivity under the inspired intellectual 
leadership of Phil Anderson. Regardless whether it has anything to say about the real life of high-{\em T}$_c$  superconductors  these deserve
attention if not only because it developed into a quite interesting and innovative branch of theoretical physics. 

The spin fluctuation theories are rooted in the  perturbative schemes  of the fermiology that was established in the 1950's.  
The Resonating Valence Bond (RVB) ideas eventually rest on the 
recognition that doped Mott insulators have to be different. This story starts with the last skirmishes of a war that is much older: the 
understanding of the Mott-insulating state itself. As I will discuss in section  2.4.3, this pursuit  started a long time ago, while it settled 
in the late 1980's when, as a byproduct of the high-{\em T}$_c$  research, a rather detailed and complete 
understanding of the real life Mott insulators was established.  Section 2.4.4 is dealing with the RVB school of thought.  This started from 
an appealing  idea by Phil Anderson
that evolved during the wild early years of high-{\em T}$_c$  in something of a mathematical science by itself. Different from the 1950's
spin fluctuations, these 'gauge theories of strongly correlated fermions'  rest conceptually on the discoveries in the 1970's in
high energy physics.  Although it still lacks both convincing experimental support
and real mathematical control, these theories surely serve well the purpose of confusing the high energy theorists. At stake is the 
meaning of the gauge principle. In fundamental physics it is  a primordial principle. The 'RVB' gauge  theories are insisting 
however that the gauge principle is governed by collectivity,
telling how large numbers of 'un-gauged' microscopic degrees of freedom co-operate in unison to generate 'gauged' collective behaviors.

Keeping the historical order, in section 2.4.5 I will focus on the subject that has been dominating the field during the last ten years or so: 
the physics of the underdoped 'pseudogap regime'. This has been so much on the foreground because it turned out that there was
much to discover using the modern experimental techniques.  It is now clear that there is a collection of exotic competing orders
at work,  including stripes, quantum liquid crystals and spontaneous diamagnetic currents. Apparently theoretical physics as we know
it does give a grip on orderly things, because  theorists did make  a big difference in this particular pursuit.  

In the last section I will discuss what may be the beginning of the history of the future of high-{\em T}$_c$ superconductivity. This revolves around the
idea that the quantum physics of the mysterious superconductors is in the grip of the overarching symmetry principle of scale invariance.
In  analogy with the classical critical state, this 'quantum criticality' is believed to be caused by a phase transition happening at zero temperature,
driven by quantum fluctuations. Given the increased understanding of the competing orders at work in the pseudogap phase this notion
is at present becoming quite popular in the cuprates. In the context of the heavy fermions it has been however obvious for
a long time that their superconductivity is closely related to quantum phase transitions and at stake is whether these share the essential
physics with the cuprates. This is the stronghold for the agnostic theorists because nowhere else it is so obvious that we are lacking 
mathematical machinery that works as in the context of quantum critical 'fermionic' metals.  I will end this chronicle on an optimistic tone
by expressing my hope that the string theorists might actually have such machinery in the offering. 
  
\section{Pushing BCS to the limits:  the spin fluctuation superglue.}

The BCS theory is a  remarkable achievement. A constant factor in the debate during the last twenty five years has
been whether it is at all possible to get around the essence of BCS to explain superconductivity at a high temperature. 
Its central wheel is the Cooper instability which in turn rests on the physics of nearly free fermions. A first condition is
that the normal state closely approaches the Fermi-liquid fixed point near the superconducting transition temperature. 
A second condition is that the net residual quasiparticle interactions  turn attractive in the appropriate pairing channel.
At this instance the Fermi-surface exerts it power, by singling out the pairing channel as uniquely singular.  Feeding a generic fermion 
system with a very strong attractive interaction that overwhelms the Fermi-energy would just cause a clumping of the particles 
since in (nearly) classical systems there is no special  stability associated with two particle bound 
states\footnote{Although it is usually worked under the rug, this problem hunts ideas involving local pairing, like the bipolarons.}. 
The remainder of BCS is simple.  Cooper pairs  are bosons and the Cooper pairs Bose condense
at the moment they form, of course modulo the possibility of 'dangerous' thermal order parameter fluctuations. 

The question is whether the normal state is sufficiently like a  Fermi-liquid  for the BCS mechanism to apply directly. 
This has been dividing the minds more than anything
else -- the 'radical' theories discussed in the next sections depart from the assertion that the normal state is something
else than a Fermi liquid.  Assuming the governance of the Fermi-liquid,   the only issue is  then 
to  explain the origin of the attractive interactions. This is far from trivial since electrons actually repel each other rather strongly
on the microscopic scale through the Coulomb interactions. Much of the hard work of diminishing these bare interactions 
can be ascribed to the magic of the Fermi-liquid. The ubiquitous nature of the  quasiparticle  gas is much more of a mystery 
than is often realized. The particular limiting case where on the microscopic scale the interactions are substantially smaller
than the bare Fermi energy is nowadays well understood, although this involves the fanciful 'Polchinski-Shankar' functional 
renormalization group. By principle one has to keep track of an infinity of coupling constants 'spanning up' the Fermi-surface, 
when integrating out short distance degrees of freedom. In the renormalization process, one finds that besides the 'harmless' marginality 
of the Landau F functions, one has only to watch a possible (marginal) relevancy in the Cooper channel.\footnote{For a fanciful recent
discussion see S.A. Parameswaran, R. Shankar and S.L. Sondhi, arXiv:1008.2492.} The bottom line is that with the exception of the BCS superconductivity
the Fermi-liquid is generically (i.e., away from special nesting conditions) stable as an infrared fixed point.   

The problem is however that the systems of interest
in the present era are invariably of the kind where  the microscopic interaction scale exceeds the bare kinetic energy by order(s) of magnitude. One
can no longer rely on the perturbative renormalization group and it is just  an experimental fact that Fermi-liquids are nevertheless ubiquitous. 
 A classic example is $^3$He. At microscopic distances this is like a quantum version 
of a dense, highly correlated  Van der Waals liquid. However, upon cooling it down to milli-kelvins one finds a long wavelength
physics corresponding with an impeccable Fermi-gas of quasiparticles.  The miracle is that the 'hard Helium balls' have become 
effectively completely transparent relative to each other, only communicating via the Pauli principle, while the effects of interactions
is that these 'quasi Helium atoms'  are three times heavier than the real atoms. I learned from 
Robert Schrieffer  that in the 1960's-70's  theorists tried hard to explain this wonder, not getting anywhere.\footnote{Remarkably, as a very 
recent development it appears  that the quantum-fields gravity holographic dualities of string theory   shed a new light on this old problem:
the Fermi liquid is related to the gravitational stability of special black holes with 'fermion hair', see section 2.4.5.}
Fermi-liquids are unreasonably resilient and this is perhaps the best reason to take the 'conservative' theories of high-{\em T}$_c$  superconductivity
serious.

 The next question becomes: how to turn the residual weak quasiparticle repulsions of the Fermi-liquid into serious attractive interactions?
The classic BCS answer is of course phonons. The exchange of bosons that are not governed by the gauge principle will generically give
rise to an induced attractive interaction. The phonons are just perfect: they are external to the electron system since they originate in the
lattice, their interactions with the quasiparticles emerge naturally and they have the added benefit of causing retarded interactions.  
Phonon energies are tiny compared to the Fermi energy of conventional metals and the associated 'Migdal parameter' can be exploited to simplify
the perturbation theory into a virtually exact 'resummed' Migdal-Eliashberg theory. 

This control over the classic phonon-driven BCS theory explains the sociological earthquake that rumbled through the physics community
following the Bednorz-M\"{u}ller discovery of superconductivity at 34K in $La_{2-x}Ba_xCuO_4$ in 1986. Back then it was believed that
phonons could not be responsible for a {\em T}$_c$ exceeding 30 K and  a consensus emerged immediately that something new was
at work. In hindsight this is  actually a bit embarrassing.  It is a number game and numbers are dangerous. The strongest claim rested on
stability considerations\footnote{M.L. Cohen and P.W. Anderson, in {\em Superconductivity in d- and f-band Metals}, ed. D.H. Douglass, AIP, 
New York (1972), p. 17.} that were subsequently challenged by theorists of Ginzburg's high-{\em T}$_c$ superconductivity group at the Lebedev 
institute\footnote{O.V. Dolgov, D.A. Kirzhnits and E.G. Maksimov, Rev. Mod. Phys. 53 (1981) 81.}. In fact, in 2001 superconductivity at 
40 K was discovered in the simple $MgB_2$ system. In no time a consensus emerged that the superconductivity is phonon driven, with
 a number of material details (high phonon frequencies, multi-bands) conspiring to optimize the phonon mechanism,
while the phonons responsible for the superconductivity are unrelated to the factors rendering the crystal to be stable. This
development gave also credibility to the claim that such phonon driven superconductivity can be quantitatively studied using LDA type band structure
theory. Computing $T_c$'s from first principles is very demanding and has only become possible recently by the increase of computer power. 
$MgB_2$ and a number of other cases have lend credibility to the case that these numbers can be trusted.  This support the 'computational'
case that 40K need not be the limit of conventional  phonon driven superconductivity\footnote{See, e.g., Marvin Cohen's talk at the 2009 KITP
superconductivity conference: http:$//$online.kitp.ucsb.edu$/$online$/$highertc09$/$cohen$/$}.

Despite these caveats I am not aware of any expert  believing  that conventional phonons can be responsible for the superconductivity in 
cuprates or  iron pnictides.  Perhaps the best reason is that the extreme electron-phonon coupling required to explain a $T_c$ of
150K should somehow reveal itself in band structure calculations, while these just indicate quite moderate couplings. This is not to 
say that phonons do not play any role. In fact, there are striking experimental indications that especially at very low doping the electron-phonon
coupling is so strong that some variety of small polaron physics is at work. This appears to invoke polar couplings that are ignored in the
LDA calculations where implicitly metallic screening is assumed. It cannot be excluded that even at optimal doping such a 'poor metal' polaronic physics
is at work and claims that bipolarons are behind the superconductivity cannot be dismissed
 beforehand.\footnote{A.S. Alexandrov and N.F. Mott , {\em Polarons and Bipolarons}, World Scientific, Singapore (1995).}

The main stream thinks however differently. Staying within the "BCS paradigm"  the quest to explain the high-{\em T}$_c$'s  turns into the issue whether
one can identify non-phonon stuff that can take the role of the phonons, mediating very strong attractive interactions.  The only option on the table
is to look for 'glue' that  emerges from the interacting electron system itself. There are quite a number of options here, like the excitonic mechanism
first proposed by Little, emphasizing special units in the crystal structure  characterized  by a high electronic
 polarizability.\footnote{This idea revived in the context of pnictides: see G.A. Sawatzky {\em et al.}, Eur. Phys. Lett. 86 (2009)174409.} 
 Another old idea is the  Kohn-Luttinger overscreening  effect in 2D systems\footnote{W. Kohn and J.M. Luttinger,
Phys. Rev. Lett.  15 (1965) 524.} and the closely related idea of the plasmon taking the role of glue. However, by far the most popular view has been
up to the present day to point at spin fluctuations. The basic idea dates back to Berk and Schrieffer in the 1960's who aimed at explaning why metals 
that are close to a transition to an itinerant ferromagnetic state usually do not superconduct\footnote{N.F. Berk and J.R. Schrieffer, Phys. Rev. Lett. 17
 (1966) 433;
an important follow up is S. Doniach and S. Engelsberg, Phys. Rev. Lett. 17 (1966) 750, focussing on the renormalization of the Fermi liquid.}. 
They introduced a particular approximate way to compute matters that forms the conceptual core of the myriad of improved spin fluctuation theories
that have been developed since then. 

A first hurdle is that in the electron system at microscopic distances there is no such thing as a system of spins that
has a separate existence of the electrons -- the electrons just carry around the spins themselves. The 'spin glue' has therefore itself to be a highly
cooperative phenomenon emerging at 'large' distances in the interacting electron system. Upon approaching a transition to a magnetic state in an
itinerant metal one does however expect precursors of the magnetism in the paramagnetic state. 
These are in the form  of short lived magnetic fluctuations that become increasingly better 
defined upon approaching the magnetic transition where they turn into real magnons -- the paramagnons. Using time dependent mean field theory
(RPA) the paramagnons can be computed perturbatively assuming weak interactions. Assuming in addition residual interactions between these spin fluctuations
and the conduction electrons, their propagators can be just feeded into the Migdal-Eliasberg formalism to take the role of phonons. 

The particular 
benefit of this approach is that it is quite predictive with regard to the pairing symmetries. For the case of ferromagnetic fluctuations Schrieffer and Berk
showed that these generically cause repulsions in the s-wave channel explaining why nearly ferromagnetic metals usually do not superconduct.
In the early 1970's  superfluidity  in fermionic $^3$He was discovered. It was already known that the normal state is characterized by a strongly enhanced 
magnetic susceptibility  indicating that the Helium Fermi-liquid is on the verge of becoming ferromagnetic. Given that there are no phonons around, 
the idea that this superfluidity  had dealings with paramagnons was immediately realized. Dealing with phonons, the induced attractive 
interaction surely favors a spin singlet order parameter, while in good metals the effective coupling is localized in space and thereby momentum independent,
favoring s-wave superconductivity. However, ferromagnetic fluctuations favor spin-triplet pairing. This in turn implies that the pairs need to be in an orbital state
with uneven angular momentum, having the added benefit that the real space pair wave function has to have a node at the origin. The 
fermions in the pair thereby  avoid each other automatically such that the hard core microscopic repulsions are circumvented. 
The bottom line is that although there might be a net repulsion between the quasiparticles this might well turn out to sum up 
into an attractive interaction in a higher angular momentum channel. 

The superfluidity in $^3$He is of the triplet variety  and a discussion doing justice to the richness of 
this subject deserves a separate chapter in a book like this.\footnote{The classic, hard to improve treatise is A.J. Leggett, Rev. Mod. Phys.  47 (1995) 331.} 
With regard to the microscopic pairing mechanism a worthwhile lesson  is the use of  the Berk-Schrieffer logic  by Anderson
and Brinkman to get with great effect  the details of the A-phase  right.\footnote{P.W. Anderson and W.F. Brinkman, Phys.Rev.Lett. 30 (1973) 1108.}
The gross message is that $^3$He demonstrates that an interacting fermion system as dominated by microscopic  repulsive interactions 
can eventually turn into a superconductor, with the pairing mechanism having undoubtedly dealings with the spin fluctuations.  However, the 
Anderson-Brinkman result is just showing that the spin fluctuations contribute to the attractive interactions, stabilizing a particular superconducting
state, but it does not prove that the RPA style spin fluctuations do all the pairing work explaining in detail why $T_c$ is what it is. For instance, much attention 
was paid as well to the role of the attractive tail of the microscopic Van der Waals interaction. Another example of this lack of microscopic understanding 
is the question whether the mass enhancements
of the $^3$He Fermi liquid can  be accounted for by the coupling to the ferromagnetic spin fluctuations or whether they are related to the proximity
of the crystallization transition\footnote{D. Vollhardt, Rev. Mod. Phys. 56 (1984) 99; this in turn rested on the discovery by Maurice Rice 
and Bill Brinkman of the description of the Mott transition in terms of Gutzwiller projected wave functions: W.F. Brinkman and T.M. Rice, Phys. Rev. B2
(1970) 4302. These ideas were in turn inspirational in the development of the RVB type theories of section 2.4.4.}.  It appears that the quest 
to explain the nature of the residual attractive force responsible for the pairing in $^3$He ended in a stalemate. Many factors can contribute and it is just
a number game.  Given that there are no controlled mathematical methods that tell us how to connect the strongly interacting microscopy to the 
weakly interacting infrared, it is in fact plainly impossible to address this subtle 
quantitative matter with the available theoretical technologies. With an eye on the remainder, one should be aware that $^3$He is in a way  a best case
scenario for glue ideas, since the normal state is a well developed Fermi-liquid, with a 
superfluidity that is clearly in the weak coupling limit, while complications like (lattice) Umklapp and  phonons are absent. 

The notion that the superconductivity  of electron systems in solids could be caused by intrinsic electronic interactions effects got on the main stage
for the first time by the discovery of the heavy fermion superconductors in the late 1970's and 1980's. The gigantic mass renormalizations in  these
systems leave no doubt that electron-electron interaction effects are dominant. In addition it became gradually clear that the superconducting order
parameters in these system are unconventional by default while  magnetism is usually around the corner. During the last twenty years a consensus
emerged that in these systems the superconductivity is closely related to the presence of a quantum phase transition where at zero temperature a
magnetically ordered state gets destroyed by quantum fluctuations. In section 2.4.6 I will take up this theme in detail, including a discussion 
of the Hertz type theories that are closely related  to  the spin fluctuations.  

The spin fluctuation idea acquired directly prominency in the early history of cuprate high-{\em T}$_c$ superconductivity. Obviously an agent
different from phonons causing the superconductivity was at work, while in the frenzy of the late 1980's it became in no time clear that cuprate 
superconductivity emerged from doping an antiferromagnet. In addition, both inelastic neutron scattering- and NMR experiments gave 
away  that the metals at superconducting doping concentrations show the signatures of pronounced spin fluctuations. The spin-fluctuation-glue
idea was lying on the shelf and it was predictably taken up by part of the community. The most visible advocates were  Douglas Scalapino 
and David Pines, representing somewhat complementary views that are in 2010 still discernible. Pines embodied the 'bottom-up'
phenomenological approach, analyzing in great detail experimental information on the spin fluctuations, aiming for
a quantitative phenomenological description of the superconductivity. Scalapino approached it from a more theoretical angle, trying to cast 
the (incomplete) information from quantum Monte Carlo computations and so forth in a diagrammatic framework taylored according to the 
Berk-Schrieffer framework and improvements there off like the  "fluctuation-exchange" (FLEX) perturbative theory.     

In fact, the spin fluctuation advocates faced a heavy battle in the early history of high-{\em T}$_c$  superconductivity. The predictive power with regard to the 
pairing symmetry was their stronghold. In the presence of Umklapp associated
with the scattering of the electrons against the crystal lattice the pairing symmetry story becomes 
more interesting and specific. The lattice warps the electron dispersions and the Fermi-surface  shape with the ramification that on the level of RPA 
calculations the spin fluctuations tend to get quite structured in momentum space. The cuprate fermi-surface computed by LDA bandstructure 
(and confirmed by photoemission measurements) shows 'nesting' features that promotes a strong enhancement of spin fluctuations at large
momenta.  This in turn implies a strong momentum dependence of the pairing interactions mediated by these fluctuations that strongly 
favor  a $d_{x^2-y^2}$ superconducting order parameter. In the 1987-1990 era it was however taken for granted by the community at large
that cuprate superconductivity was s-wave. In hinsight these strong believes are not easy to explain on rational grounds: the data were of a 
poor quality and it appears to be rooted instead in a conservative sociological reflex  -- even d-wave was too risky. 
After a hard fight, that lasted for a couple of years Scalapino and Pines {\em et al.}  were proven right with their d-wave claim by the phase sensitive 
measurements of Tsui, Kirtley, van Harlingen and coworkers discussed elsewhere in this book.   
This is remarkable and I find it respectable that this success has motivated a stubborn
attitude in the spin fluctuation community: there has to be something to the notion given that it got the d-wave right. However, after it became clear 
that the pairing symmetry is d-wave in no time it turned out that this is also the natural ground state in for instance the RVB context (section 2.4.4),
and even invoking the spontaneous current phases of Varma as discussed in section 2.4.5. Although d-wave superconductivity
is rather unnatural dealing with conventional phonons, the consensus has emerged that it is just too generic dealing with electronic mechanisms for it to
be regarded as conclusive evidence for any specific mechanism.   

At  this moment in time the idea of a spin-fluctuation 'superglue' is very much alive -- counting heads in the community 
it might be the most influential hypothesis all together. Part of this appeal is surely associated with the fact that it poses a crisp challenge 
to the experimentalists: the superglue is like a holy grail, and one just has to dig with the modern experimental machinery of condensed matter
physics into the electron matter to isolate it. Fact of the matter is however that after twenty years of concerted effort in the cuprates 
the 'evidence' is as indirect as it has  been all along.  The state of the art is perhaps best represented by a tour de force  that was published in 
2009.\footnote{T. Dahm {\em et al.}, Nature Physics 5 (2009) 217.} Empirical information obtained from direct measurements of the spin fluctuations
by neutron scattering and the electron self energy effects picked up in photoemission is combined with the results from
numerical calculations, arriving at the claim that  the spin fluctuation glue can fully account for the superconductivity. But this work also reveals
the weaknesses of this 'paradigm'. The evidences are highly quantitative, while it departs from oversimplified toy models where 
one nevertheless looses mathematical control because of the intermediate strength of the couplings involved.  Equally convincing cases
have appeared arguing that the self energy kinks in photoemission are entirely due to rather weak electron-phonon coupling.  
The magnetic resonance that dominates the spin fluctuation spectrum in optimally doped cuprates can be interpreted 
rather as an effect than as the cause of the superconductivity. 
Perhaps the most devastating  counter evidence available at present for such an interpretation 
is the staggering 'electronic complexity' of the underdoped pseudogap regime as discussed 
in section 2.4.5. The simple fermiology  'RPA view' on this competing order appears as a gross oversimplification of this reality. 

The advent of the pnictide superconductors appeared initially as a shot in the arm for the spin fluctuation idea. The LDA calculations indicate nesting
properties of the Fermi-surface pockets that are beneficial for the RPA paramagnons, while they are consistent with the "$(\pi, 0)$" magnetism 
found in the parent compounds. The case was made vigorously that since the parent systems are not Mott insulating it has to be that the iron 
superconductors are quite itinerant and typical candidates for spin fluctuation superconductivity. Last but not least,  a peculiar "$s_{\pm}$" pairing 
symmetry was predicted based on the LDA-RPA-Eliashberg "trinity".\footnote{This appears to have been a close race, with winners: I.I. Mazin {\em et al.}, 
Phys. Rev. Lett.  101 (2008) 057003, being cited  already 500 times.} It is now widely believed that this symmetry is indeed realized in at least some
of these systems, although definitive evidence is still missing. 

Much of the early experimental activity in this young field was aimed at trying to find evidence for the spin-fluctuation superglue.
Remarkably, when the moment seemed fit to cry victory the building started to crumble. Data appeared indicating that 
 unexpected 'nematic'  phenomena are happening in the underdoped pnictides that seem to parallel the pseudogap complexities of the cuprates.
 In addition,  the 
pnictide metals at optimal doping have weird properties like a linear resistivity that again suggests that these have uncanny commonalities with
the mystery stuff of the cuprates. This is a very fresh development and it is much too early to say where it will land. However,  I do expect that 
it is  beneficial for the newcomers in the pnictide field to be aware of the very long and contentious history of the  spin fluctuation glue idea.

\section{The legacy of Philip W. Anderson (I): Mottness.}

When in some future the real history of $T_c$ can be written, the odds are high that Philip Anderson will have a legendary role to play.
The influential Andersonian tradition in high-{\em T}$_c$ superconductivity  revolves around the assertion that the ground rules of metal physics as laid down
in the 1950's  do not apply to the electrons in cuprates, for the reason that high-{\em T}$_c$  superconductivity emerges from a Mott insulator that is 
doped. This is in stark contrast with the spin-fluctuation school of thought as discussed in the previous section. This departs from the notion 
that one can get away with the idea that the normal state is in essence a Fermi-liquid. Although shaken hard by the powerful glue
bosons, the quasiparticles forming a Fermi-gas are eventually ruling the waves and it is just a matter of 
sorting  out the Feynman graphs governing the perturbations of the quasiparticle gas.   When high-{\em T}$_c$  was 
discovered in 1986 this  was taken as a self-evident truism by nearly everybody. However, Anderson published one of 
the first papers ever on the subject and this is still the most influential paper in the history of the subject introducing the 'resonating valence bond'
(RVB) idea\footnote{P.W. Anderson, Science 235 (1987) 1196.} that will be the focus point of the next section.  This section is a
preliminary:  the success story of nailing down the specifics of the parent cuprate Mott insulator in the late 1980's.   

 Mott insulators are in fact much simpler than band insulators: they are
just the incarnation of rush hour traffic  in the world of electrons. The Coulomb repulsion dominates with the effect that 
electrons have to avoid each other.  When only one electron fits every unit cell, while there are as many unit cells as electrons all 'electron 
traffic comes to a complete stand still'. Doping has the effect of making some room for motions -- it is the hbar version of stop and go traffic.  The first
fact that got settled in high-{\em T}$_c$  superconductivity was the Mott insulating nature of the undoped cuprate parents. 

This matter actually jump started  my own career. I was quite lucky having 
George Sawatzky as thesis supervisor. In the early 1980's  photoemission was still quite young and George had the vision to use this
technique to focus in on the classic problem of the electronic structure of insulating transition metal salts. This was also the era that the local density
approximation (LDA) band structure description of the electronic structure was at the height of its success, having demonstrated that it  delivered a remarkably 
accurate description of the band structures of conventional metals and insulators. Unaware of its contentious history, the claim was  made 
by band structure theorists in 1984 that  straightforward spin dependent  LDA did describe the electronic structures of 3d transition metal oxides   
correctly. This triggered a ferocious response by Anderson. I have a vivid recollection reading a short paper he wrote that never got published, 
written with a literary quality that is rare in the physics literature: "Williams et al. arrive at the laughable contention that cobalt oxide is a metal because 
their band structure calculation is telling so."  

Apparently this response was rooted in developments in the 1950's. Slater might be viewed as the 
father of modern band structure theory. With his $X-\alpha$ method he had outlined the idea of effective exchange-correlation mean-field potentials
and one can view the later work of Kohn and so forth as an a-posteriori formal justification of the basic notion. Apparently Slater believed that his mean-field
potentials were the last word and it is rumored that Slater gave the young Anderson a very hard time. 

In this era Mott and van Vleck,  Anderson's intellectual fathers, had discovered the basics of what we nowadays understand as 'strongly correlated 
electron physics'.  This revolved initially around the notion that the electron physics in transition metal salts is dominated by strong, local electron
repulsions.  Although Hubbard can be credited to be the first who wrote down the Hubbard model explicitly in 1964, Mott introduced the notion of Mott
insulator in 1956, while Van Vleck already had in the 1940's a clear notion of the local Coulomb interaction ($U$) competing with the 
hopping $t$\footnote{According
to Phil Anderson, Van Vleck used to call it the UT model, a typical side reference to the University Theater, Harvard Square ...}. Anderson 
himself has a big stake in this early development in the form of his 1959 work on the superexchange mechanism for the spin interactions 
in Mott insulators\footnote{P.W. Anderson, Phys. Rev.  115 (1959) 2.}, as well as his Anderson impurity problem (awarded with the 
Nobel prize), representing landmarks of the physics  of strongly correlated electron systems.   

In the period where the band structure/fermiology view was turning into the main stream the solid state physics community was not particularly receptive to 
the Mott-insulator bad news.  When this old struggle revived in 1984, we were making in Groningen good progress,  finding interesting signals of the 
'Mottness'  at work in 
transition metal salts in the form of satellites and so forth in various electron spectroscopies. Together with Jim Allen, Sawatzky realized that this new information
could be used to make a difference in the old debate and this culminated in the 'Zaanen-Sawatzky-Allen' theory for the electronic structure of Mott-insulating
salts.  We explained how  the essence of the toy models designed by Anderson and Hubbard could be combined with the complexities of the electronic 
structure of the real salts. In the first half of the 1980's nobody seemed interested, but this changed
drastically when high-{\em T}$_c$ superconductivity was discovered. Cuprates were an important part of the 
ZSA ploy as archetypical 'charge transfer insulators' 
where the holes of the p-type cuprates are actually formed from oxygen states. My thesis work was to a 
degree rediscovered by others: Emery with his 'Emery model' and Zhang and Rice with their 'singlet' (Sawatzky's idea of starting with 3d
impurities).  Rather quickly this converged into a main stream view, insisting that one should be able to get away in the doped systems 
 with the minimal t-J model, much in the spirit of Anderson's RVB philosophy. The marked exception was Chandra Varma. He was convinced  
 that 'chemistry matters',  that one has to worry about holes being on oxygen. After a lonesome but resilient search of some twenty years this triumphed recently,  in the form of Varma's intra-unit cell spontaneous currents, as will be discussed in section 2.4.5.

In the years that followed a massive effort ensued aimed at understanding the parent Mott insulators. 
The experimental- and theoretical technologies were just lying ready to be used to completely
clarify their physics. It appears that  this resulted in a seemingly complete understanding of Mott-insulators. 
We have learned how to modify LDA so that it can cope quantitatively  with the real life
Mott insulators through the LDA+U and LDA+DMFT functionals. A detailed understanding has been achieved of the physics of an isolated carrier moving through the antiferromagnet resting on linear spin-wave theory and the self-consistent Born approximation, 
with the recent addition that in the insulators the polar electron-phonon couplings are
actually quite important. The study of Mott-insulators has turned in the mean time into a form of standard science in material-science laboratories,
but unfortunately this has not made much of a difference with regard to the still highly mysterious physics of Mott-insulators turning into metals and 
superconductors when doped.       

\section{The legacy of Philip W. Anderson (II): Resonating Valence Bonds and its descendants.}

Upon learning in 1986  that  high-{\em T}$_c$  superconductivity did occur in a doped copper-oxide,  Anderson directly arrived at the hypothesis  that 
the phenomena should be rooted in the 'strangeness'  of the electron states subjected to the "Mottness".  His RVB paper 
heralded a remarkably intense and creative period in condensed matter theory.  The 'spins and holes' ideas that are still at the center of the 
thinking regarding unconventional mechanisms were laid down in a high pace in the years 1987-1988, under the intellectual leadership of  Anderson.
The emphasis is  here on 'ideas': it actually amounts to inspired guesswork, not disciplined by controlled mathematics, while until the present day 
empirical evidence for 'RVB-ish' physics is quite thin. With the exception of the notion that Mottness is changing the ground rules, my impression is that
even Anderson  will admit that the problem of high-{\em T}$_c$  superconductivity is still wide open. However, regardless whether they apply literally to cuprate 
superconductors, the ideas that emerged in 1987/1988 are embodying a very interesting theoretical framework that deserves to be realized by nature
somewhere.

The essence of the Mott-insulator is that the electrons that should form a high density degenerate Fermi-gas, turn instead into a system of spins due
the domination of the local Coulomb repulsions.   This is simple to understand, but the consequences are remarkable. The free Fermi-gas is in first instance governed 
by Fermi-Dirac statistics, rooted in the fact that electrons are indistinguishable fermions. Due to the interactions the electrons localize, thereby becoming 
distinguishable particles,  and all what is left behind are their spins that live as distinguishable degrees of freedom in tensor product space.  Upon doping,
the moving holes release 'here and there' the indistinguishable nature of the electrons. In the doped Mott insulator the electrons are most of the time 
distinguishable spins, but once in a while when a hole passes by they remember that they are actually fermions. At stake is that 'Mottness'  interferes with  the 
fundamental rules of quantum statistics. This is a rather recent insight -- I learned it from Zheng-Yu Weng, and it appears  that it is not quite appreciated yet
by the community at large. Whatever it means,  I find it myself  a particularly helpful point of departure to appreciate why RVB and so forth makes 
sense.\footnote{see J. Zaanen and B. Overbosch, arXiv:0911.4070. See also P. Phillips. Rev. Mod. Phys. Coll.  82 (2010) 1719 for a complimentary perspective on Mottness. } 

The point of departure of Anderson's seminal RVB paper is the idea that the pairing of electrons is already imprinted in the spin system formed in the Mott-insulator.
The common phenomenon in these Heisenberg spin systems is that the spins order in simple antiferromagnets. After the dust settled in 1987/1988 this turned
out also to be the case in the high-{\em T}$_c$  parents. However, in two dimensions and dealing with small $S=1/2$ spins there is a potentiality that the
quantum  fluctuations   become so strong that some quantum spin liquid is formed. Anderson envisaged such a liquid, actually inspired by old ideas of Pauling. Configurations from pairs of spins forming singlets ("valence bonds") are formed, and the spin liquid 
corresponds with coherent superpositions ("resonating") of all possible tilings of the lattice with such singlet configurations. Upon doping,
the electrons forming the singlet pairs become mobile, and the valence bonds turn into Cooper pairs that condense in a superconductor. 

The idea is simple and appealing but the hard  part is to demonstrate that the doped spin system is actually dominated by these pair-wise singlet configurations. 
After all these years definitive evidence for such 
a state is still missing. There is no controlled mathematical- or computational procedure available for the doped Mott-insulator problem, as rooted
in the statistical troubles alluded to in the above ('fermion signs').  At the same time, there is no experimental machinery available that can measure
the RVB 'entanglement' directly and the evidences that are quoted to support RVB type physics are invariably indirect -- the pseudogaps, d-wave superconductivity  and so forth can be as well explained within e.g. the conventional spin fluctuation view. 

The 'idea revolution' that ensued in the period 1987/1988 was driven by the desire to pack mathematical meat on the conceptual bones of Anderson's RVB
idea. Anderson took a first step in this direction in his 1987 Science paper by introducing the idea of a particular kind of variational wave-functions.   This dates back  to the helium era where Jastrow introduced the idea of Ansatz wavefunctions that are derived from the free Fermi/Bose gas but where one wires in
that the particles avoid each other locally due to strong local repulsions.\footnote{R. Jastrow, Phys. Rev. 98 (1955) 1479.}  Gutzwiller suggested a special
form of such Ansatz wavefunctions , just amounting to a free fermion ground state where the double occupied states are 
projected out.\footnote{M.C. Gutzwiller, Phys. Rev. Lett. 10 (1963) 159.} These were in 1972 applied by Maurice Rice and Bill Brinkman,
in the context of the Mott metal-insulator problem. Anderson pointed out in his RVB paper that projecting out double occupancy from  a free BCS 
ground state wavefunction (instead of the Fermi gas) yields a state which is just like his envisaged RVB superconductor. To do this well one has to evaluate 
matters numerically, and this variational Monte-Carlo approach turned into a main stream with hundreds of papers published. 

Despite strong claims, this numerical activity has not been decisive up to the present day. The Gutzwiller style variational approaches suffer intrinsically from 
the 'garbage in-garbage out' problem. One wires in 
through the Ansatz a particular prejudice regarding the physics and it is hard to judge from the outcomes whether this was a good idea, given the well established 
wisdom that ground state energies and so forth are not very sensitive measures of the correctness of the wave function. In fact, the failure of the computational methods at large can be viewed as the best evidence for the profundity of the doped Mott insulator problem. All numerical methods 
characterized by  a real measure of  control show that they eventually loose this control. Quantum Monte Carlo and the closely related high 
temperature expansions crash before temperature is low enough for matters to become truly significant because of the sign problem. 
The density matrix renormalization group fails in two (and higher) dimensions by its inability to keep track of the quantum entanglements. 
Dynamical mean field theory (DMFT) is based on the uncontrolled  ad-hoc assumption  that the effects of   electron correlation effects are purely local 
and in this regard it is not that much different from a primitive Gutzwiller Ansatz. Right now perhaps the most promising
approach is the 'dynamical cluster approximation' as designed by Jarrell  that can be viewed either as a finite cluster Quantum Monte Carlo 
with  cleverly designed effective medium boundary conditions, or either as a finite range extension of the DMFT. At the least this approach has 
delivered a real surprise in the form of a phase separation type quantum phase transition, happening in the Hubbard model at 
intermediate coupling.\footnote{see E. Khatami {\em et al.},  Phys. Rev. B81 (2010) 201101.} 

The theoretical pursuit triggered by the RVB idea eventually settled in what is called the
'gauge theories of strongly correlated electron systems'. Several motives that were around got combined into a new view
on the collective behaviors that at least in principle can exist in strongly interacting quantum matter. Perhaps the most important motive is the 
notion of quantum number fractionalization:  the idea that the strongly interacting, highly collective state of matter might carry excitations that behave like
weakly interacting particles/elementary excitations carrying  however 'fractions' of the quantum numbers of the elementary particles. In the context of the 
RVB school of thought this turned specifically into the notion that the electron quasiparticle of Fermi-liquid theory 'falls apart' in particles that carry separately
spin 1/2 ('spinons') and  electrical charge ('holons').  The fractionalization 'phenomenon' was first identified in the 1970's  by Jackiw and Rebby in the 
mathematical study of field theories in one space dimensions (Thirring-, Sine-Gordon models) while it acquired great fame in condensed matter physics
when it was rediscovered by Su, Schrieffer and Heeger in the  simple,  physical setting of the excitations  of 
polyacetylene.\footnote{see the classic review: A.J. Heeger, S. Kivelson, J.R. Schrieffer and W.P. Su, Rev. Mod. Phys. 60 (1988) 781.}

This is just a dimerized 
(double bond-single bond) chain of carbon atoms and 'SSH' recognized that a kink/domain wall in this dimerization pattern carries an electronic bound
state that is formed from half a conduction band- and half a valence band state. The consequence is that it corresponds with a pure S=1/2 excitation  carrying 
no charge ('spinon'). When one binds a carrier to this soliton it turns into a spin-less excitation carrying the electron charge ('holon'). Kivelson, Rokshar 
and Sethna recognized 
that a 'short range' RVB state formed from nearest-neighbor singlet pairs (described by 'quantum dimer' models)  is for the purpose of fractionalization
just a two dimensional extension of the SSH physics.\footnote{S.A. Kivelson, D.S. Rokhsar and J.P. Sethna, Phys. Rev. B35 (1987) 8865.} 
As long as the dimers form a liquid it is simple to see that isolated spins and -holes are sharply defined 
topological excitations that carry either spin 1/2 or charge. Fradkin, Kivelson and others soon thereafter managed to link the quantum dimer models to 
gauge theory.\footnote{See Eduardo Fradkin's  book {\em Field theories of Condensed Matter Systems}, Addison-Wesley (1991).}
They showed that it is dual to a version of compact quantum electrodynamics in two space dimensions, linking the fractionalization to 
the confinement-deconfinement phenomenon, as known from e.g. quantum chromodynamics. 

However, compared to the high energy physics context the
perspective is warped in a way that continuous to confuse high energy theorists. The lore in fundamental physics is to consider the quarks as the 
fundamental particles and due to the anti-screening property of the gluon fields, these are confined at low energy in baryons. The rules precisely reverse
in the quantum dimer model: the electron is the fundamental object, but it only exists in the confining state of the theory. The spinons and the holons
have the status of the quarks, but they appear as highly collective excitations rooted in the global properties of the 'topologically ordered' RVB liquid. 
These only become real when the ground state/vacuum of the gauge theory is deconfining. Therefore,  deconfinement is a profound emergence phenomenon
while the confining state is just about no-brainer microscopic  'chemistry' surviving up to the longest distances. This interesting motive is shared by all 
theories dealing with the 'emergent gauge principle' in condensed matter physics. 

The quantum dimer problem on the square lattice actually died in a rather interesting way. 
Compact pure gauge electrodynamics is always confining in 2+1D. In the 'dual' quantum dimer model context this translates into the fact  that at least on a square
lattice the ground states always correspond with valence bond crystals where the valence bonds just stack in a regular manner breaking the lattice translations. 
This has the consequence  that spinons and holons are confined, since they are tied together by domain wall 'strings' in the valence bond crystals. 
Much later Moessner and Sondhi discovered a loophole, by their
demonstration that the dimer model of the 2D triangular lattice does support a deconfining state.\footnote{R. Moessner and S.L. Sondhi, Phys. Rev. Lett. 
86 (2001) 1881; see also A. Kitaev's honeycomb model, Ann. Phys. 321, (2006) 2.}  This is perhaps still the best proof of principle that fractionalization
as linked to the notion of deconfinement in gauge theories does make sense, at least theoretically.

Another, at first view unrelated way, that the gauge principle enters was recognized by Baskaran and Anderson, now directly associated with the very 
nature of Mottness itself. This relates directly to the 'stay at home' principle that emerges at low energy in the Mott-insulator. The Mott-localization implies
that in fact the {\em local} number operator is sharply quantized: in the Mott-insulator one can count precisely  the (integer) number of electrons within
a finite localization volume. Baskaran and Anderson argued that one can work with free fermions as long as one couples these to a compact U(1) gauge 
field that in turn imposes the local conservation of charge.\footnote{G. Baskaran and P.W. Anderson, Phys. Rev. B37 (1988) 580.} 
Affleck and Marston took this one step further by noticing that the Heisenberg type spin Hamiltonian is in 
turn also invariant under charge conjugation (it doesn't matter whether one describes spins in terms of electrons or holes), showing that together with
the 'stay at home' gauge the overall symmetry of the gauge fields in the Mott insulator is SU(2).\footnote{I. Affleck {\em et al.}, Phys. Rev. B38 (1988) 745.} 

The final crucial motive that was required for the mathematical underpinning of the gauge theories for doped Mott-insulators is the idea 
of the large $N$ limit. $N$ refers here to the number of degrees of freedom in a field theory. The notion that one can organize matters in terms of a
perturbation expansion where $1/N$ acts as the small quantity dates back to the hay days of non-abelian Yang-Mills theory in the 1970's. The Dutch
theorist 't Hooft realized  that the diagrammatic perturbation theory of the pure gauge $SU(N)$ Yang-Mills theory drastically simplifies in the
limit that $N \rightarrow \infty$, since one only has to consider the so-called planar diagrams. When the gauge coupling is strong (the interesting- and physically
relevant case) one still has to re-sum an infinity of diagrams but these now form dense two dimensional "nets" in space time that are topologically trivial
(no handles).  This motive evolved in a crucial conceptual pillar underpinning the gauge-gravity dualities of modern string theory: these planar diagram 
networks can be viewed as string worldsheets that are topologically trivial, with the ramification that
 the large $N$ limit of the gauge theory behaves like string theory at small
string couplings. 

The way that large $N$ enters in the context of strongly interacting electron systems is actually very different, obviously
so because there is no such thing as fundamental non-abelian gauge bosons at the microscopic scale. This involved a leap of imagination 
by  Ramakrishnan and Sur,\footnote{T.V. Ramakrishnan and K. Sur, Phys. Rev. B26 (1982) 1798.}  who showed that the large degeneracy
of the strongly  interacting 4f atomic states in the valence-fluctuating/heavy fermion intermetallics could be exploited for the purposes of a large N expansion.
It soon became clear that the Kondo- and Anderson impurity problems drastically simplify in the limit of high orbital degeneracy. A next crucial step was taken 
by S. Barnes, followed soon thereafter by   Newns and Read with the discovery that 'slave' mean field theories control the physics at large 
$N$.\footnote{See D.M. Newns and N. Read, Adv. Phys. 36 (1987) 799 for an account of this early era of slave theories. Piers Coleman 
deserves  special mention for his work on the impurity models, apparently be the father of the idea of slave particles: Phys. Rev. B 29 (1984) 3035. } 

This construction can be considered as the central wheel in the 'modern' formulation of the emergent gauge 
principle in strongly interacting fermion matter. It demonstrates the potentiality of gauge symmetry- and dynamics  as a controlling 
principle for the emergent highly collective physics at long distances and times, in a manner that is completely different from the way the gauge principle 
is understood 'as coming from god' in high energy physics.  At first sight the construction seems absurd. One starts out with for instance an Anderson 
lattice model describing a lattice of strongly interacting f-shell electrons hybridizing with a non-interacting valence band, such that one of  the two low energy 
valence states of the f-shell has a large orbital degeneracy $N$. One then asserts that the f-electron field operators can be written as products of 
creation/annihilation  operators  of two particles,  one carrying the spin- ('spinon') and the other the charge ('holon') of this electron. The electron is a 
fermion and therefore one has to make a choice how to divide the quantum statistics over the slave particles. One can either associate a bosonic
statistic to the holon such that the spinons are fermions ('slave fermions'), or the other way around  ('slave bosons') , while pending the choice one ends
up with quite different theories. Surely these slave particles are unreal: they correspond with redundant (gauge) degrees of freedom that can be 
removed by imposing gauge invariance invoking explicit gauge fields. But then one discovers that in the large $N$ limit the local constraints that 
tie the spinons and holons together turn into simple global chemical potentials, which is the same statement that the gauge couplings disappear.
The consequence is that the effective gauge theory is in a deconfining regime such that the spinons and holons become as real as quarks above the confinement transition!

One is not done yet because the resulting theory in terms of 'physical' spinons and holons has no free part: it only contains two particle (spinon-spinon,
spinon-holon) interaction terms. But these can be handled in terms of mean-field theory, introducing single boson- (Bose condensation) and bilinear 
fermion vacuum expectation values. In the original context of Anderson lattice models this was used to explain the emergence of the heavy electron
bands. However, soon after the RVB ideas of Anderson had sunk in a variety of theorists realized more or less independently that this large $N$ 
slave theory route seemed quite promising with regard to explaining high-{\em T}$_c$  superconductivity when applied  to the appropriate large $N$
generalizations of the t-J model. In case that the fermion statistics is attached to the spinons these mean field theories are quite like the Gutzwiller type theories 
for RVB states,  while they also include the fractionalized excitations suggested by the quantum dimer theories. The spin vacuum is build up from pair singlets with  a hard-wired d-wave type coherence, described in terms of a BCS-like mean field structure at 'infinite' coupling. The doped holes turn into holons,
that Bose condense directly giving rise to a 'holon superfluid' with a superfluid density that increases with the number of doped carriers. At the same
time, the spinon BCS state has a $T_c$ that decreases with doping. From gauge invariance it is easy to understand that one has a physical superconductivity
only when both holons and spinons are condensed, explaining the doping dependence of the superconductivity. 
Above all, it  becomes trivial to explain  why $T_c$ is high because the scale is set by the large superexchange interaction $J$. The underdoped  state 
at higher temperatures where the spinons are condensed, but the holons are still normal is suggestive of the pseudo-gap regime, while it can also
be argued that the (overdoped) state where the spinons form a metal while the holons are condensed should eventually correspond with a Fermi-liquid.

The benefit slave theories  in the large $N$ limit surely lies in the sense of mathematical control while the physical implications are 
remarkable.\footnote{For a recent in depth review see, P.A. Lee, N. Nagaosa and X.-G. Wen, Rev. Mod. Phys. 78 (2006) 17.}
The headache is however that $N$ in the physical case is only two ( spin degeneracy) and it is highly questionably whether the large $N$ limit has anything
to say about the real system. For any finite $N$  the gauge couplings in the bare theory jump to infinite since  Maxwell terms for the gauge fields 
are non-existent. This would mean that one lands in a strongly confining regime, where the spinons and holons bind in the original electrons such that nothing is 
achieved. But one can then argue that short distance perturbative corrections involving the spinons and holons will generate dynamically  a gauge kinetic
energy.  This might then stabilize a deconfining state, but the next problem is that one has to find reasons that in the two space dimensions of the cuprates
the monopoles associated with the compactness of the gauge field can be kept out of the vacuum. This gave rise to much debate over the last twenty 
years and a consensus emerged that in principle even in two dimensions the deconfining state has a chance to exist in principle.  

The slave particle gauge theory is up to the present day highly contentious as an explanation for
superconductivity at a high temperature, and not only because of the lack of real mathematical control in the physically relevant small $N$ regime. 
The greater problem is that despite many attempts this theory has failed to produce striking insights
and predictions for experiment. It offers suggestive explanations for  a 
number of observations, but these can also be explained in a multitude of different ways (like the superconducting dome, the pseudogap,
and so forth). In order to be taken seriously,  a theory that  departs as radically from established wisdoms  as these gauge theories, it should produce
'smoking gun' predictions that can be tested by experiment.  After more than twenty years, such hard empirical evidence is still not available.

Another matter is that this development in high-{\em T}$_c$  context has inspired a  revival of the field of quantum magnetism. 
Pure spin systems tend to order "classically"  in the form of (anti)ferromagnets. When the microscopic spins have a small magnitude (like the S=1/2 spins of
the cuprates) the quantum fluctuations around this classical order can be quite serious. The challenge is to find out whether quantum spin systems 
exist that do not order in a conventional magnet but instead have a 'quantum spin liquid' ground state. This is  a natural arena of the  slave particle gauge
theories discussed in the above paragraphs.  There are a number of indications that the slave theories are landing on the real axis in this context.
An early success is the spinon theory in bosonic incarnation by Arovas and Auerbach.\footnote{D.P. Arovas and A. Auerbach, Phys. Rev. B38 (1988) 361.} 
 This representation is especially suited to track the physics of  quantum spin systems that fail 'at the last minute' to turn into classical order:
it turned out to be very closely related to the renormalization group for the quantum non linear sigma model discussed in the last section. A next success
was the large N slave fermion spinon theory in the $CP_N$ incarnation by Sachdev and Read\footnote{N. Read and S. Sachdev,
Phys. Rev. B42 (1990) 4568.}, predicting the valence bond crystals in two
space dimensions. It appears that in the mean time  this state is now well established in the $J_1-J_2$ frustrated Heisenberg 
spin systems\footnote{This got much attention in the entertaining context of 'deconfined quantum criticality', T. Senthil {\em et al.}, Science 
303 (2004) 5663.},  while the case
is compelling that this valence bond 'crystallization' is playing an important role in the stripe physics discussed in the next section. The key question is however 
whether {\em compressible} quantum spin liquids exists, systems that show no manifest classical order while their excitation spectrum is gapless. 
Believing  the fermion slave theories such states should be generic, corresponding with states where the fermions form effective Fermi-liquid
systems characterized either by large Fermi-surfaces or states that look like unconventional superconductors with massless nodal excitations. 
In fact, in this context the most fanciful mathematical considerations were formulated, in the form of the classification of spin  liquids in terms of the
projective symmetry groups by X.-G. Wen.\footnote{This was introduced in: X.G. Wen, Phys. Rev. B65 (2002) 165113.} 
This work exposes more than anything else the unusual nature of the symmetry at work in the 'slave worlds'.
The essence of Wen's construction is the recognition  that the gauge volume of the slave theories is not spanned by separate gauge fields but instead by 
gauge equivalent slave mean field ground states. Wen derives the precise symmetry algebra's for this 'projective symmetry', and shows how to derive
the elementary excitations in Goldstone theorem style, coming up with a zoo of compressible 'fermion like' spin liquids. 

Quite recently an organic system was discovered in Japan where the empirical evidences are quite serious for the existence of such a 'spinon Fermi-liquid'.
This system  forms a geometrically frustrated  two dimensional lattice of organic molecules containing one electron per site, 
forming a Mott insulator that is very close to 
the transition to a metal. Applying a small pressure it undergoes a  first order transition into a metallic  state that suggestively turns into an unconventional superconductor
at low temperatures. The insulating magnet does not show any hint of long range order down to the lowest temperatures while several 
properties are quite suggestive of a "spinon" Fermi-liquid type of physics.\footnote{See P.A. Lee, Science 321 (2008) 1306.}

\section{Theory that works: the competing orders of the pseudogap regime.}

In the cuprates the emphasis shifted in the last decade or so to the investigation of the underdoped pseudogap phase.  The reason is that 
high Tc superconductivity is a very empirical branch of physics and there is much to explore in the pseudo gap phase using the modern machines
of condensed matter physics. This has developed into a kind of embarrassment of the riches. It is now clear
that competing order is at work, in tune with the idea of a quantum phase transition at optimal doping (next section), but the evidence
points  actually at a whole collection of exotic ordering phenomena. Not so long ago there was much disagreement among the different sub-communities
focussed on particular orders,   but a consensus emerged that these are all  real. The present state of affairs was 
anticipated a number of years ago by Bob Laughlin, who described this situation in his popular
book\footnote{R.B. Laughlin, {\em A different Universe: Reinventing Physics from the Bottom Down}, Basic books (2005).} as the 'dark side of emergence',
arguing that the strongly interacting electron matter of the cuprates is complex to a degree that approaches biology.  The ramification would then be 
that theoretical physics is side-tracked dealing with such complexity.  Although Laughlin's prophecy regarding the  pseudogap phase
turned out to be quite accurate,  I do find his view overly pessimistic. In the first place, when one heats up the pseudogap "stuff' one reenters 
the "marginal Fermi-liquid"  normal state at some higher temperature. As I will discuss in the next section, this is by itself a monument of simplicity 
that deserves to be  explained by equations of an Einstein-Dirac quality.
The  significance of  the pseudo-gap 'complexity' then lies in the fact that apparently  the  simple quantum critical metal of the cuprates  carries the {\em potentiality} to "fall apart" in a variety of exotic ordered states.  
Andy Mackenzie gives it a wonderful engineering twist: the stuff that makes up the quantum critical metals of the 
next section corresponds with construction material that makes possible to build states of quantum matter that cannot be constructed from stable states like 
the Fermi liquid. 

Laughlin gets it really wrong with regard to the role of theorists. Although my guild has failed miserably with regard to impacting on the understanding 
of the disordered quantum liquid states, theory has been remarkably successful in guiding the experimentalists in discovering the exotic orders at work in
the pseudogap phase. Apparently the mathematical machinery that is lying on the theoretical physics shelf is particularly geared towards understanding
order in the sense of symmetry breaking. 

Among the phenomena that have been identified in the pseudogap phase,  only the notion of low superfluid density, strongly phase
fluctuating superconductivity was in principle understood before 1987.  The naive version of this is the idea of 'preformed pairs' but the reality in cuprates
is better captured by Laughlin's metaphorical 'gossamer superconductivity'.\footnote{Given that the concept of phase fluctuating 
superconductivity is quite conventional, the literature is large. It  starts with several groups in the late 1980's that are largely ignored because 
they were ahead of the fashion.  At least counting citations, V.J. Emery and S.A. Kivelson, Nature 374 (1995) 434  has been perhaps most influential. 
Laughlin stresses the proximity to the Mott insulator as a perspective to understand the 'thin superconductor': R.B. Laughin, Phil. Mag. B86 (2006) 
1165.} The phase fluctuations of the superconductor are caused by the 
struggle for domination of the superconductivity with the competing states.  The proximate Mott-insulator that hangs like a big thundercloud over the pseudogap regime is in this regard a major factor (the 'gossamer' idea). 

The  other 'exotic' pseudogap orders that have been identified in the mean time  were invariably 
not known in 1987.  These  were first discovered theoretically, motivating experimentalists to look for them in less obvious corners.  These gained 
general acceptance only after much controversy: testimony for the novelty involved in the discoveries.  In historical order this is about  the stripes, 
the quantum liquid crystal orders and  the intra-unit cell  spontaneous diamagnetic  currents.      

The stripes were the first to be discovered and they have at present the status of best established among the exotic pseudogap orders. They came 
into existence  on some Friday afternoon in early October 1987, when I was fooling around with a computer code that produced a very unexpected
output.\footnote{For the record, this work was presented for the first time in a seminar in december 1987 at the ETH Z{\"u}rich, and 
 in a large oral session at the Interlaken M2S meeting in Februari 1988. The paper was submitted in March 1988 to Physical Review Letters and 
 after 1.5 years it was transferred to Physical Review B while the original submission date was removed: 
 J. Zaanen and O. Gunnarsson, Phys. Rev. B  40 (1989) 7391. I kept the files with referee reports -- it is a hilarious read. A couple of months 
 after submission I got an e-mail by Machida explaining that he had seen our preprint while they had independently discovered the same motive. 
 All other papers in this early era came much later. Especially, the late Heinz Schulz approached me after my Interlaken talk explaining that he found 
 it very interesting, planning to take up the theme by himself. Apparently, back in Paris he had forgotten where he got the idea.}  This landed on the real 
 axis by the observation of static stripes with neutron scattering in the so-called 214 LTT cuprates by Tranquada {\em et al.} in 
 1995.\footnote{J.M. Tranquada {\em et al.}, Nature 375 (1995) 561.} while it took until this century for it to become fashionable. 
 At another occasion I will give a detailed account of this long  history. The bottom line is that I got myself eventually lavishly rewarded for my resilience,
 but I also learned to be quite cynical regarding the agility and open mindedness of the relevant part of the physics community.  
 
What are the stripes about? When it started in the 1987  it was perhaps in first instance a symbol for the idea that the electron matter in cuprates could be 
more organized than in conventional metals. The  resistance against the idea that the electron stuff  could be 'inhomogeneous'  is just  
rooted in the way hard condensed matter physicists are trained. Much emphasis is put on the physics of Fermi-
and Bose gasses that make much sense dealing with the kinetic energy dominated electron matter in conventional metals. However, dealing 
with doped Mott-insulator, rush hour traffic is a more useful metaphor: when the insulator is like the complete jam, stripes are like the dynamical patterns 
that occur when the electron traffic enters the stop- and go regime of the lightly doped insulator\footnote{See J. Zaanen, Science 315 (2007) 1372  
for a further elaboration of this metaphor.}  Early on, there were more theorists playing around with the inhomogeneity motive, including
Lev Gorkov,  as well as Vic Emery and Steve Kivelson  who were pursuing in the late 1980's the idea of frustrated phase
separation droplets.\footnote{V.J. Emery and S.A. Kivelson,   Physica C209 (1993) 597.}

Stripes are however  a quite specific form of organization that is directly tied to the proximity of the Mott insulator.  
The carriers form lines ('rivers of charge') in the two dimensional cuprate planes, separated by Mott-insulating 'stripes' that support antiferromagnetic
order, with the specialty that the 'rivers  of charge' have a double role as domain walls in the antiferromagnetism. From the point of view of symmetry breaking these
correspond just with a conventional electron charge density wave coexisting with an incommensurate colinear antiferromagnetic order ('spin density wave'). 
However, the microscopic physics  is quite special for the doped Mott insulator. The main contribution of Olle Gunnarsson to the first paper
was his vision that the stripes should be viewed as the electronic version of a  {\em discommensuration lattice} -- we borrowed the
designation 'stripes' from the naming of such states as they occur in classical incommensurate systems. Hubbard  and t-J models
are just crude toy models, introduced to capture some essence of the physics. Viewed generally, Mott insulators are electron crystals that are
commensurate with the underlying ionic lattice, and the Mott gap is nothing else than a 'commensurate pinning energy' causing a gap in the acoustic phonon 
spectrum of the electron crystal. Upon doping, the lattice constant of the electron crystal becomes incommensurate relative to the ion lattice and since 
the pinning is strong the mismatch gets concentrated in small areas (discommensurations, the charge stripes) while elsewhere  the electron crystal 
stays commensurate (the Mott insulating domains). A difference with the classical incommensurate systems is the spin degeneracy. This is lifted 
by the short range quantum fluctuations turning the spin system into an incommensurate antiferromagnet.    My 1987 computation was just rooted 
in Hartree-Fock ('classical saddle point', 'old fashioned mean field theory', whatever) and dealing with classical orders (like antiferromagnets, 
electron crystals) the power of this method is at least in qualitative regards remarkable. It got the stripes quite right.

Stripes are now well known to be ubiquitous in doped Mott-insulators. Usually (nickelates, manganites) they submit however to the Hartree-Fock
stability rule that they should keep the doped Mott insulator insulating. The cuprates are in fact the exceptions since they can be made metallic. Static stripes 
seem exceptional in cuprates: they occur at the crystal surfaces (the scanning tunneling microscopy measurements) while only in very special (214-LTT) 
cuprates they seem to exist in the bulk, where they are very bad for the superconductivity.  So much is clear that in the cuprates
stripes are shaken hard by quantum fluctuations. The density matrix renormalization group computations
by Steve White and Doug Scalapino in the mid 1990's were quite revealing in this regard.\footnote{S.R. White and D.J. Scalapino,
Phys. Rev. Lett. 80 (1998) 1272.}  Different from the 'deeply classical' Hartree-Fock stripes, 
the stripes in the t-J model seem to be governed by a RVB like organization of the spin system, such that they can be viewed  as in essence a crystallization of 
valence bond pairs.\footnote{As first suggested by S. Sachdev and N. Read, Int. J. Mod. Phys B 5 (1991) 219:
this appears to be quite consistent with the $Cu-O-Cu$ 'valence bond' units comprising the stripes seen by the STS 
'Mott maps' by the Davis group.} Quantum physics also seems to rule the stripes in the long time limit. 
  Resting on the information obtained on the dynamical spin fluctuations by inelastic neutron scattering the 
case has been forcefully made\footnote{Our Leiden group was in this regard (too) far ahead of the time, investing much energy in the problem of fluctuating
stripes in the early 1990's: See J. Zaanen, M.L. Horbach and W. van Saarloos, Phys. Rev. B53 (1996) 8671 (1996) and the review
 J. Zaanen {\em et al.} Phil. Mag. B 81 (2001)1485. See also S.A. Kivelson {\em et al.}, Rev. Mod. Phys. 75 (2003) 1201 and  for a
 recent review see: M.  Vojta, Adv. Phys.  58 (2009) 699.} that the quantum liquid in underdoped cuprates looks like a stripe state that has
fallen prey to what the quantum field theorists call 'dynamical mass generation'. These dynamical stripes might behave as static stripes at short distances
but  going to longer times the {\em collective} quantum fluctuations associated with the order increase, to get out of
hand at a relatively long time, beyond which one is dealing with a uniform quantum liquid. 

It has proven difficult to get a handle on these dynamical stripes. However, inspired by this question a next great idea was born. Static stripes should 
in first instance be seen as an electron crystal. Resting on the analogy with the 'partial' melting of classical crystals due to thermal fluctuations, 
Steve Kivelson, Eduardo Fradkin and Vic Emery introduced in 1997 the idea\footnote{S.A. Kivelson, E. Fradkin, and V.J. Emery, Nature 393 (1997) 550.}
 that, given the severe quantum fluctuations, the stripe crystal in the 
cuprates could perhaps quantum melt step by step.  This would then yield  a hierarchy of quantum liquid crystal states: the quantum smectic 
(crystal in one direction,  quantum liquid in the others)   and the quantum nematic (a quantum liquid breaking spontaneously rotational symmetry).

I was immediately enchanted by the idea. Realizing that this was potentially a highly fertile territory, I made it into a main focus of my own research,
but as far as I am aware this was for a long time  the only  substantive effort outside the Kivelson-Fradkin group. The lack of interest in this obviously interesting 
pursuit was quite frustrating but the show got eventually saved again by the experimentalists.  Right now the quantum liquid crystals are 
turning into a fashionable research area. In the cuprates the first evidences were produced by the 
inelastic neutron scattering work by the group of Keimer, followed by the very recent work of Tranquada {\em et al.} on the 'sliding phase' physics in
the LTT cuprates, and especially the Nernst effect anisotropy discovered by Taillefer's group and the peculiar 'hidden' orientational symmetry
breaking in the STM stripes of Davis et al. This is supplemented by the particularly neat quantum nematic found at the metamagnetic quantum 
critical point in a ruthenate by the group of Mackenzie , as well as the very recent evidence for the  famous hidden order in $URu_2Si_2$  to be
of this kind. Perhaps the most spectacular evidence was presented in 2010, in the form of the extreme transport anisotropies in 
detwinned underdoped pnictides by the Fisher group.

I find the greatest star among the exotic pseudogap orders to be the spontaneous diamagnetic current phase that started with the theoretical work by
Chandra Varma.\footnote{C.M. Varma, Phys. Rev. Lett. 83 (1999) 3538.} 
This opinion is motivated by the fact that this type of order is completely detached from anything that was known before 1987. 
The essence of the phenomenon is that below the pseudogap temperature  diamagnetic currents start to flow spontaneously
around plaquettes in the copper-oxide lattice. 

The idea that something of this kind can happen dates back to the wild early years of high-{\em T}$_c$  superconductivity, where it arose in the context 
of the gauge theories discussed in  section 2.4.4.  In  their study of the slave mean field theories controlled by  gauge symmetry, 
Ian Affleck and Brad Marston\footnote{I. Affleck and J.B. Marston, Phys. Rev. B37 (1988) 3774.} observed that besides the spinon d-wave superconductor there was another stable mean field state. In this 
 'staggered flux phase' diamagnetic currents occur spontaneously, forming a staggered pattern on the plaquettes of the square lattice.
In the context of the gauge theories this flux phase has a very special status:  in Wen's modern projected symmetry language,  the d-wave superconductor and 
the flux phase span up the gauge volume in the sense that one can {\em locally} rotate one in the other without physical ramifications for the spin liquid. Upon doping
this gauge invariance is lifted and the flux phase turns into a real phase that takes the role of competitor of the d-wave superconductor. 
A main result of the work by Patrick Lee and coworkers on the $SU(2)$ formulation is the prediction  that one should
find collective  modes in the d-wave superconductor that correspond with exciting infinitesimal flux phase modes.\footnote{P.A. Lee and N. Nagaosa, 
Phys. Rev. B68 (2003) 024516.} 

This idea was recycled in 1999 by Chetan Nayak\footnote{C. Nayak,  Phys. Rev. B62 (2000) 4880; this was quite effectively advertised by
S. Chakravarty {\em et al.}, Phys. Rev. B63 (2001) 094503.}
 showing that staggered flux phases have a generic significance also in a more conventional (Hartree-Fock) setting. It is well known that under certain fine tuning 
conditions one can store the particle-particle Cooper pairs and particle-hole charge density wave in a single (global)  $SU(2)$ multiplet (Anderson's 
pseudo-spins), where the superconductivity is associated with the XY plane and the CDW with the z-direction of the pseudo-spin internal space. In this way one 
can show that the $SU(2)$ partner of a s-wave superconductor is a conventional charge density wave. Nayak observed that a d-wave superconductor
'rotates' in a similar guise into a staggered flux phase. This motivated the name d-density wave (d-DW), and it gave further impetus to the idea that
as such it would be a
natural candidate for the 'hidden' competing pseudo-gap order. However, despite a concerted experimental effort  this idea failed to generate any 
empirical support.

Varma actually departed from a quite different perspective. From the very beginning of the cuprate odyssee  he had been in the grip of the idea that
one has to look in more detail into the chemistry of the cuprate planes, taking a quite dissident stance in a climate dominated by the divinity
of the simple t-J model.  Varma departed from the wisdom that the holes are 
physically located on the oxygen ions, while he observed that in the presence of 'excitonic' Cu-O nearest neighbor 
Coulomb interactions one can run into instabilities when  the doping increases. Much later, he realized that such microscopic physics opens 
up the possibility for d-DW type currents to develop inside
the Cu-O unit cell, forming patterns involving the various $O-Cu-O$ intra-unit cell triangular plaquettes. The specialty of such a
phase is that it  breaks only time reversal symmetry, with the ramification that one has to know where to look for it in experiment in order
to find it.  The first indications for the presence of such a phase in the pseudogap regime were derived from ARPES measurements,
but the data analysis turned out to be a contentious affair.  This landed on its feet in 2005 when Philippe Bourges {\em et al.} found direct 
evidence for such intra-unit cell 'antiferromagnetism' breaking time reversal symmetry using rather straightforward spin-polarized neutron scattering. Around 
the same time complimentary evidences were found by Kapitulnik {\em  et al.} employing highly sensitive optical dichroism measurements. 

I was among the first to see this paper -- as 'Mr. Stripes' I has perhaps  something to loose, but I was immediately convinced 
that Varma's currents are real.  Given the persuasiveness of the evidence, and the fact that this was a genuine new type of ordering phenomenon 
I took it for granted  that this would make headlines by the next day. But reality was quite different. Again the brick wall of hard headed conservatism
was erected. The authors had a hard time to get this published and initially it was largely ignored. As with the stripes and the nematics,
the neutron scattering community once again demonstrated its
open mindedness. In a concerted effort, with a particularly prominent role of Martin Greven with his superb mercury crystals, much additional evidence
was generated.  The latest evidence is by Philippe Bourges {\em et al.}, showing that in the 214 stripe phase the stripes not only live
together with nematic order (according  to Davis et al.), but also with a short ranged Varma current order, making the case  that the pseudogap
phase deserves a name that  was recently conceived by Uchida: the monster.

Summarizing, although the pseudogap regime has still its mysterious sides (like the normal state 'Fermi-arcs', and the 'two gap' behavior) the theorists
can at the least claim that they have played a decisive role in uncovering the various exotic ordering phenomena that are taking place. Another matter
is what this history reveals regarding the sociology of this community. As I implicitly emphasized, to my perception the community operates in this regard
in a less than optimal way. Scientific skepticism is a great good and perhaps it is just part of human nature that scientific communities have to be conservative.
However, an atmosphere that would have been a bit more receptive towards exciting new scientific potentialities would have undoubtedly rendered 
the above research enterprise to be a more efficient affair.   

 \section{Quantum critical metals and superconductivity.}

A last major theme in the context of theories of unconventional superconductivity is the school of thought that revolves around
the idea that the physics of the cuprates , heavy fermion superconductors, and perhaps the  pnictides and organics revolves around
a quantum phase transition. The emphasis is on 'quantum':  the transition is happening at zero temperature and driven 
by quantum fluctuations. In a way this general idea has many fathers. It is also a prevalent theme  elsewhere in physics:
I recently learned that in string theory the notion of what is known in condensed matter as quantum criticality 
is among the fundamental organizing principles in this seemingly very different area of physics. In fact,  I will end this history reporting
on a very recent attempt to mobilize the powerful mathematics of string theory to corner the physics of high-{\em T}$_c$  superconductivity. 

Symmetry is the best friend of the theoretical physicists and the critical state (quantum or not) is in the grip of one of the most powerful
symmetries that has been identified: scale invariance. This came to flourish in the classical 
realms in the 1970-1980's by the discovery of Wilson's renormalization group theory, which is in essence just the theory that exploits the
power of the scale invariance emerging at the classical continuous phase transitions. In a way it is remarkable that it took so long before
it was realized how to exploit its full powers in the quantum realms. In high energy physics this is likely due to the pre-occupation with 
QCD. This is about the 'marginal' case with its running coupling constants, as echoed in condensed matter physics in the form of
the attention to the Kondo impurity problem and the non-abelian 'Haldane' spin chains subjected to the phenomenon of dynamical mass generation. It 
appears that in string theory conformal field theory (also beyond the 1+1D worldsheet context) came fully into focus when supersymmetric field
theories were explored. Due to the non-renormalization theorems coming with supersymmetry these are characterized by strongly interacting
unstable fixed points (in the statistical physics jargon) that span up planes in the space of coupling constants.

 In condensed matter physics
the idea of quantum phase transitions entered the main stream in 1988, by the seminal work of Sudip Chakravarty, David Nelson and 
Bert Halperin (CNH).\footnote{S. Chakravarty, D.R. Nelson and B.I. Halperin, Phys. Rev. B39 (1989) 2344; In this same era the notion of 'quantum critical 
hydrodynamics' was also fully realized in the context of DC transport phenomena both involving (fractional) quantum Hall plateau 
transitions (S.L. Sondhi {\em et al.},  Rev. Mod. Phys. 69 (1987) 315) and the superconductor-insulator 
transition in amorphous systems (M.P.A. Fisher {\em et al.}, Phys. Rev. B40 (1989) 546; {\em ibid.}, Phys. Rev. Lett. 64 (1990) 587).} 
 Bert Halperin told me that this work was
initially motivated by his conviction that Phil Anderson got it all wrong with his claim that the magnetism in  the  Mott insulating $La_2CuO_4$ had to do with
resonating valence bonds.  Bert got it right: the quantum magnetism in this system is of the quasiclassical kind, controlled by the collective quantum 
precessions of the N\'eel order parameter. Compared to many other antiferromagnets, this $S=1/2$  two dimensional Heisenberg spin system is not that 
far from the point where these precessions get completely out of hand, quantum melting the N\'eel order in a continuous quantum phase transition. 
This physics is captured by the O(3)  quantum non-linear sigma model field theory which amounts to a 'classical' Heisenberg system that is 
living in the 2+1 dimensional space time of the euclidean (thermal) path integral. Sudip Chakravarty was back then on sabbathical in Harvard 
and it appears that due to his  high energy physics training he was pivotal in the realization that this becomes really interesting when one asks 
what the proximity to the quantum phase transition means for the {\em finite temperature} physics. As far as I am aware they were the first to realize
that by raising temperature one re-enters the quantum critical 'wedge' where the low energy physics is controlled by the zero temperature quantum
physical scale invariance. They demonstrated how the spin stiffness controlling the thermal order parameter fluctuations is renormalized by the quantum 
fluctuations  as was soon thereafter spectacularly confirmed experimentally. 

In the same year the Bell Labs theorists Varma and Littlewood with coworkers had independently constructed the marginal Fermi liquid 
phenomenology for the metallic states of cuprates at optimal doping.\footnote{C. M. Varma {\em at al.}, Phys. Rev. Lett. 63 (1989) 1996. 
Undoubtedly Phil Anderson also played a key role in attracting attention to the strangeness of the normal state metal. With his 'tomographic Luttinger
liquid' (see his book) and his more recent 'hidden Gutzwiller quasiparticle gas' (Nature Physics  4 (2008) 208) he was (and is) searching for 
a truly non perturbative formulation for a state that behaves temporally in a critical way while it is still 'organized around a Fermi surface'. This contrasts strongly with
the perturbative (Hertz-Millis like) attitude of the marginal Fermi-liquid. At least on the phenomenological level his constructions are remarkably
similar to the very recent results coming from the Anti-de-Sitter/Conformal Field Theory 
 correspondence. Historically the marginal Fermi-liquid had perhaps more impact for no
other reason than its lack of pretense, to be no more than a way to memorize the strangeness in the experimental data.}  
This is actually not much of a mathematical theory, but it was quite effective 
in highlighting- and unifying the strange behaviors of these normal state metals in a single phenomenological (or 'heuristic', 'mnemonic') 
framework. Forced by the experimental circumstances Varma {\em et al.} imported the 'Planckian dissipation' (energy/temperature scaling) as a crucial 
building block of the construction, among others giving a heuristic explanation for the famous linear-in-temperature resistivity of the optimally doped metals.

 Subir Sachdev discovered soon afterward that  the classical, relaxation hydrodynamics 
inside the quantum critical wedge  is highly unusual\footnote{S. Sachdev and J. Ye, Phys. Rev. Lett. 69 (1992) 2411}. Much later I named it 'Planckian dissipation'\footnote{J. Zaanen, Nature 430 (2004) 513; 
embarasingly, I called such a state maximally viscous not realizing that viscosity is proportional to the relaxation time. The Planckian dissipator is therefore 
as close to a perfect (un-viscous) fluid as possible, as was realized in the same period in the context of string theory and the quark-gluon plasma.}:  
it is a dissipative state, controlled by Planck's constant through a 'universal' relaxation time $\tau_{\hbar} \simeq \hbar / (k_B T)$.  In my own reference frame I became  fully aware of the meaning of quantum criticality-proper at the Aspen winter conference in 1992 where 
Sachdev  presented their analysis of the NMR spin-spin and spin-lattice relaxation times in superconducting, underdoped cuprates employing the Planckian 
dissipation.\footnote{A.V. Chubukov and S. Sachdev, Phys. Rev. Lett. 71 (1993) 169.}  The implication of this work was that at least the spin
fluctuations of the optimally doped 'marginal Femi-liquid' was in the grip of the quantum scale invariance. 
This was an important moment in my career: I have been since then a devotee of the quantum critical
state, in fact as a goal by itself regardless its relevance towards superconductivity and so forth. 
During the 1990's Sachdev developed the quantum critical hydrodynamics theme further and this culminated in his quantum phase transition book.
\footnote{S. Sachdev, {\em Quantum Phase Transitions}, Cambridge Univ. Press (1999).} 

The theme was picked up in the high-{\em T}$_c$  community, since the relevancy of the notion towards the then widely accepted marginal Fermi-liquid
phenomenology was rather obvious. A difficulty was however that for quantum criticality to control the physics at optimal doping it occurs that one 
needs a quantum phase transition, and for this one needs some form of order that comes to an end in the best superconductors. Loram and Tallon
came up early on with thermodynamical evidences for such a quantum critical point but this was back then highly controversial.  As described in the
previous section in the last few years this has changed drastically with the solid empirical evidences for the 'monster' competing orders that are
simultaneously at work in the underdoped regime. It remains to be seen whether a single order parameter is responsible for the quantum critical
regime in the sense of a 'conventional' isolated quantum unstable fixed point.  A very recent idea is that the competing orders of the underdoped
regime are more an effect than a cause, reflecting instead the collapse of the Mott-ness at optimal doping which is the real driving force behind 
the quantum criticality at optimal doping.\footnote{As argued by for instance Philip Phillips, the Hubbard projections might come to an end for 
increasing doping when U is finite (Rev. Mod. Phys. 82 (2010) 1607; for closely related ideas in the heavy fermion context see C. Pepin, Phys. Rev.
Lett. 98 (2007) 206401). As we argued (arXiv:0911.4070), the big Fermi-surface in the overdoped regime might be taken as evidence that 
the Hubbard projections have disappeared in the overdoped regime (in the spirit of the considerations In Anderson's book)  while the Mott-ness
dominated underdoped regime falls prey to competing orders because of the altered 'Weng quantum statistics'. The ramification would be that
the quantum criticality is rooted in a quantum-statistics clash between the different statistics of the Mott-ness and Fermi-liquid regimes.} Altogether,
the idea that the cuprate superconductivity is rooted in the quantum critical nature of the optimally doped metal is at present again very popular 
but more work needs to be done to nail down how it precisely works. The empirical situation is in this regard more  transparent in the development
that came next: the heavy fermions.
 
This unfolded in the early 1990's with the discovery of a phletora of quantum phase transitions in the heavy fermion intermetallics,
as triggered by the seminal work of the Lonzarich group at Cambridge.\footnote{see, e.g., J. Zaanen, Science 319 (2008) 1205  and references therein.} 
These quantum phase transitions involve rather straightforward magnetic
order that is disappearing. However, on both sides of such quantum critical points (QPT) one finds Fermi-liquids that are characterized by a quasiparticle 
mass that tends to diverge at the critical point. The fermions are in one or the other way  'on the critical surface'  and perhaps the most dramatic 
empirical fact is that one invariably  finds that a 'last minute' instability sets in, centered at and 'shielding' the quantum critical point. Usually this corresponds 
with an unconventional  superconductor that takes over from a normal state showing the traits of a quantum critical metal like a linear resistivity.  
Since these systems involve  the localized f electrons of lanthanides or actinides the microscopic scales are much smaller that in the cuprates.
All along it has been realized that by 'boosting' up their microscopic energy cut-off, the physics of the heavy fermion systems would be strikingly
similar to that of the cuprates, adding credibility to the idea that also in the latter case the superconductivity is eventually rooted in the quantum 
criticality itself.

As discussed in section 2.4.2,  the idea that spin fluctuations are responsible for the cuprate superconductivity has been around since the very 
beginning. Catalyzed by the discoveries in the heavy fermion systems this merged with the quantum criticality idea, actually falling back 
on a theoretical interpretation of quantum criticality  that was introduced many years  before CNH  entered the stage. This dates back to 
the development of paramagnon theory discussed in the second section. Surely, in the context of itinerant magnetism  paramagnons become significant
when one gets close to the transition to ordered magnetism. In 1976 John Hertz reformulated this in the modern thermal field theory language,
respecting the essence of quantum criticality, but eventually ending up with a story that is operationally similar to the RPA treatment (the Moriya
paramagnon theory.)\footnote{J. Hertz, Phys. Rev. B14 (1976) 1165, with some corrections by A. Millis, Phys. Rev. B48 (1993) 7183.} 
 It starts out with the weakly interacting Fermi-gas,  the Hubbard-Stratonovich order
parameter fields are introduced, and it is asserted that the fermions can be integrated out on the RPA level, forming a heat bath
damping the order parameter. This lifts the effective dimensionality of Euclidean space time felt by the order parameter dynamics such that it lands
at- or above its upper critical dimension, implying that the fixed point is Gaussian. In the modern era it was realized that these order parameter
fluctuations 'backreact' on the fermions in the form of causing a singular interaction in the pairing channel.\footnote{For a review see Ar. Abanov, A.V. 
Chubukov and J. Schmalian, Adv. Phys. 52 (2003) 119.}  This in effect boils down  to a fancy
version of the spin fluctuation 'superglue' idea, where now the spin fluctuations appear in a quantum critical incarnation causing the superconducting
dome surrounding the quantum critical point. 

This has turned into a quite popular view both in the heavy fermion  and cuprate communities. It can even be 
correct but the problem is that the construction rests on a manifestly uncontrolled approximation scheme.  The fundamental problem are the 
fermion signs. Regardless the representation, fermions are not governed by an effective  Boltzmannian physics
in Euclidean space time due to the destructive workings of the Fermi-Dirac statistics rendering the quantum partition sum to be 
one with mathematically ill behaved 'negative probabilities'.  In dealing with a fermionic critical state one can therefore a-priori not fall back 
on the 'hidden' classical criticality, which is the secret behind the rigor of the CNH/Sachdev treatment of the bosonic problem. In the Hertz
treatment this 'fermion nightmare' is worked under the rug in the beginning, by the assumption that the fermions can just be integrated out.
The flaw is that the integration of the fermions requires them to be massive  but  the fermion sector is massless and therefore the Fermi surface 
itself is somehow part of the zero temperature critical state. In fact, it has become  clear more recently that in some
heavy fermion systems the Fermi surfaces reconstruct drastically at the quantum critical point, demonstrating directly that the fermionic degrees of freedom
are "active" part  of the criticality.\footnote{See the recent review: H. von L\"ohneisen {\em et al.}, Rev. Mod. Phys. 79 (2007) 1015.}   

With regard to the key question of why $T_c$ is as high as it is we have arrived at the present day. Although one still finds plenty of strong opinions regarding
the truth of spin fluctuations or the divinity of  RVB in the community, the 'agnost view' I referred to in the beginning has become steadily more influential 
in the community.  At least my own perspective has been changing considerably in the
last quarter century, from the desire to just explain why $T_c$ is high to the conviction that the cuprates just have been instrumental in making visible a very deep
and general problem in fundamental physics at large.  The modern code word is 'quantum matter',  the question regarding the general nature of matter as governed 
by the fundamental laws of quantum physics and the general principles of emergence. Remarkably, it appears that physics is entering an era where 
the empirical condensed matter- and the mathematical  high energy communities are merging their activities for a head on attack on this, in a way, very
new problem.  

Some twenty years ago there was an implicit but widespread  believe that we knew what matter is. 
However, high-{\em T}$_c$  superconductivity  and related phenomena have played the role of the Michelson-Morley experiment, opening our eyes 
for the fact that we actually have no clue and that there is plenty of room for surprises. I already alluded a number of times to the 'fermion signs' , the 
'negative probabilities' that spoil the mathematics that we can handle, which also show up in various guises in problems that are not governed by
fermion statistics (like frustrated spins, bosons when time reversal is broken, any non-equilibrium quantum-field problem). The bottom line is that 
these 'quantum signs' are ubiquitous and in their presence we have no mathematical tools for the description of matter that really works. The exception
are the incompressible states (like the fractional quantum Hall states, topological insulators) where topology is helping us out, but for compressible 
states it is just guess work: we just know from experiment that Fermi-liquids and the 'Hartree Fock states' (including those of section 2.4.5) are part 
of nature's portfolio of forms of matter. 

The problem of fermionic quantum criticality is rather interesting from this viewpoint. Scale invariance is a powerful symmetry and it should somehow 
render the general sign problem to become more easy. However, departing from the fifty year old established mathematical technology  one runs immediately
into a deep problem: to have perturbative control one needs the fact that the Fermi-liquid has a scale (the Fermi-energy). What to think of a fermionic
matter without Fermi energy, and how to think about a Fermi-surface when there are no quasiparticles? In the Hertz approach and so forth these questions
are worked in the rug since these constructions are departing implicitly from the stability of the  Fermi-liquid. 
We need  obviously a  mathematical weaponry to address these issues which is not lying on a shelf somewhere.

With the reservation that it is very fresh,  it might be that in this regard 2010 will end in the history books as the year that it started to happen.
During the last forty years,  string theorists have been working hard developing some very 
new mathematical machines. Although their performance as mere mathematical constructs is breath taking, it has been quite frustrating that  
these string-machines do not seem to have an obvious application to explain observable phenomena in nature.  But this is now changing:  it appears 
that the most powerful machine of string theory (the Anti-de-Sitter/Conformal Field Theory correspondence) is just tailored to compute the properties of states  of sign-full quantum matter.
The way this works is at first sight outrageous, but it becomes very beautiful when one get used to the idea. It departs
from the notion that in a special sense black holes behave as material objects when combined with quantum physics. They carry entropy and so forth,
and in their simplest incarnation they just behave as black body radiators giving off thermalized Gaussian fields. The modern string theoretical version 
has turned this into a 'generalized particle wave' duality, insisting in a surprisingly precise mathematical way that such black hole worlds in d+1 dimensional
space times are in a dual sense equivalent to strongly interacting forms of quantum matter living in a boring flat d-dimensional 
space time -- the stuffs that one finds in the earthly material science and cold atom labs! 

Right now the string theorists are sorting out collections of colorful 'hairy black holes' that by themselves represent a leap forward
in quantum gravity. These in turn render a description of quantum matter that extends way beyond the fifties paradigm. It started with the recognition
in 2007 by Sachdev, Son and coworkers\footnote{C.P. Herzog {\em et al.}, Phys. Rev. D75 (2007) 085020, see also J. Zaanen, Nature
448 (2007) 1000 for a light perspective.}
 that the correspondence encodes the 'Planckian dissipation' hydrodynamics associated with
the finite temperature quantum critical state in the condensed matter systems. Since last year matters this development speeded up, 
by the discovery that   Fermi-liquids emerge 'out of the 
blue' from a strongly interacting critical ultraviolet, while their stability is encoded by a quantum-mechanical 'Dirac hair' hanging some distance way
from the black hole horizon.\footnote{M. Cubrovic, J. Zaanen and K. Schalm, Science 325 (2009) 439.} However, such 
'Dirac hair' can be 'shaved away'  and the resulting 'extremal' Reissner-Nordstrom black-hole   codes for an emergent quantum critical fermion state.
This is breathtaking: it describes an infrared that has a sole temporal scale invariance, while the 'matching' with the bare ultraviolet fermion generates 
a singularity structure in momentum space that corresponds with the generalization of the Fermi-surface to the non-Fermi-liquid 
realms\footnote{H. Liu, J. McGreevy and D. Vegh, arXiv:0903.2477, T. Faulkner {\em et al.}, Science 329 (2010) 1043; see also S. Sachdev, arXiv:1006.3794 for inspirational ideas 
of how this relates to the matters discussed in section 2.4.4.}  This is quite like the marginal Fermi-liquid, which was introduced on mere empirical grounds!

Last but not least, the 'hairy black hole' pursuit started in 2008 by the '$H^3$' discovery\footnote{S.A. Hartnoll, C. Herzog and G. Horowitz, Phys. Rev. Lett.
101 (2008) 031601.} of the holographic encoding of the superconductor in terms of 'scalar hair'.  This tells us a fascinating story of a 'glueless' BCS like
superconductivity, where the superconducting instability is driven by a kind of 'perfect quantum frustration' of the non-superconducting quantum critical state (of the
'bald' extremal kind), as manifested by the fact that the 'naked' quantum critical point would carry a zero temperature entropy that is 'eaten' by the 
superconductor. This resonates with the idea that has been around in especially  the heavy fermion community for quite some time: 
were it not for the intervention 
of the 'last minute' superconductivity one would have to deal with a Fermi-liquid acquiring an infinite quasiparticle mass right at the critical point, 
representing of course such a  quantum frustration. The end of the present history is that right now the first attempts are undertaken to nail down 
this quantum frustration in the experimental laboratories.\footnote{see J. Zaanen, Nature 462 (2009) 15.}    

I might be overly optimistic but I have to admit that the developments discussed in the
 last paragraphs have been a source of inspiration to do my best to formulate the present 'chronicles
of the dark ages of quantum matter'.  In case that the black hole dreams land on the real axis, the professional historians will be quite curious regarding the hairy details of the prehistory of this scientific revolution. When this fails, it doesn't matter since at some point 
in the future  likely a very young physicist will scribble equations on a black board of a quality that in no time a general consensus will emerge 
that the mystery of high-{\em T}$_c$  superconductivity is solved. Given our present understanding of the depth of the mystery there is no doubt in my  mind
that the insights gained from these equations should be at least as good as the black holes!

\end{document}